\documentclass[journal=ancac3,manuscript=article]{achemso}

 \usepackage{svg}
 \usepackage{amsmath,bm}
 \usepackage{mathrsfs}
 \usepackage{amsfonts}
 \usepackage{float}
 \usepackage{setspace} 
 \usepackage{graphicx}
 \usepackage{epstopdf}
 \usepackage{dcolumn}
 \usepackage{amsmath}
 \usepackage{epsfig}
 \usepackage{indentfirst}
 \usepackage{psfrag}
 \usepackage{subfigure}
 \usepackage{amssymb}
 \usepackage{color}
 \usepackage{units} 
\usepackage{physics}
\usepackage{dcolumn}
\usepackage{bm}
\usepackage{bbold}
 \usepackage{dcolumn}
 \usepackage{natbib}
 \usepackage{wrapfig}
 \usepackage{siunitx}
\usepackage[backref=none,bookmarksnumbered=true,bookmarks=true,bookmarksopen=true,colorlinks=true,citecolor=blue,linkcolor=blue,anchorcolor=green,urlcolor=blue,unicode=false]{hyperref}

\usepackage{ulem}[normalem] 

\makeatletter
\newcommand\colorsout[1]{\bgroup \markoverwith{\textcolor{#1}{\rule[0.5ex]{2pt}{0.4pt}}}\ULon}

\makeatother

\author{Katerina Vaxevani}
  \affiliation{CIC nanoGUNE-BRTA, 20018 Donostia-San Sebasti\'an, Spain}

\author{Jingcheng Li}
  \affiliation{CIC nanoGUNE-BRTA, 20018 Donostia-San Sebasti\'an, Spain}
  \alsoaffiliation{School of Physics, Sun Yat-sen University, Guangzhou 510275, China}

 \author{Stefano Trivini}
  \affiliation{CIC nanoGUNE-BRTA, 20018 Donostia-San Sebasti\'an, Spain}
  
\author{Jon Ortuzar}  
  \affiliation{CIC nanoGUNE-BRTA, 20018 Donostia-San Sebasti\'an, Spain}
  
  \author{Danilo Longo}  
  \affiliation{CIC nanoGUNE-BRTA, 20018 Donostia-San Sebasti\'an, Spain}
  
\author{Dongfei Wang}  
  \affiliation{CIC nanoGUNE-BRTA, 20018 Donostia-San Sebasti\'an, Spain}
  
\author{Jose Ignacio Pascual}
  \affiliation{CIC nanoGUNE-BRTA, 20018 Donostia-San Sebasti\'an, Spain}
\alsoaffiliation{Ikerbasque, Basque Foundation for Science, 48013 Bilbao, Spain}\email{ji.pascual@nanogune.eu}

 \title{Extending the spin excitation lifetime of a magnetic molecule on a proximitized superconductor}

\begin{document}

\begin{abstract}
\setstretch{1.2}
Magnetic molecules deposited on surfaces are a promising platform to individually address and manipulate spins. Long spin excitation lifetimes are necessary to utilize them in quantum information processing and data storage. Normally, coupling of the molecular spin with the conduction electrons of metallic surfaces causes fast relaxation of spin excitations into the ground state. However, the presence of superconducting paring effects in the substrate can protect the excited spin from decaying. In this work, we show that a proximity-induced superconducting gold film can sustain spin excitations of a FeTPP-Cl molecule for more than  80~\si{\nano\second}. This long value was determined by studying inelastic spin excitations of the S=5/2 multiplet of FeTPP-Cl on Au films over V(100) using scanning tunneling spectroscopy. The  spin lifetime decreases with increasing film thickness, in apparent connection with the gradual gap-closing of a pair of de Gennes-Saint James resonances found inside the superconducting gap. Our results elucidate the use of proximitized gold electrodes for addressing quantum spins on surfaces, envisioning new routes for tuning the value of their spin lifetime. 
\end{abstract}
\date{\today}
\maketitle

Magnetic molecules have been under the spotlight for new quantum applications, such as quantum information processing, sensing  and data storage.\cite{gaita-arino_molecular_2019} Magnetic molecules can be tailored to behave as quantum bits (qubits), given that they exhibit long coherence (T$_2$) and spin-relaxation times (T$_1$) \cite{Ardavan_2007,Moreno-Pineda2018}. The former is one major prerequisite for engineering efficient molecular qubits  \cite{gatteschi_molecular_2006,affronte_molecular_2009, sieklucka_molecular_nodate} as it describes the lifetime of their superposition state, while the latter gives the upper limit to T$_2$. Incorporating molecular spins on solid-state platforms is the next step towards creating electrically addressable quantum devices. However, the interaction with metallic electrodes opens fast decaying channels that damp the spin dynamics. Thus, there is a need of finding metallic substrates for magnetic molecules that do not alter their spin dynamics while allowing electrical access.

Previous measurements on single transition metal atoms on surfaces reported spin excitation lifetimes up to a few hundreds picoseconds \cite{khajetoorians_itinerant_2011, khajetoorians_spin_2013,loth_controlling_2010,hermenau_long_2018}. Inserting a Cu$_2$N decoupling layer over the metallic surface can 
enlarge  spin-relaxation lifetime to nanosecond time scales \cite{loth_measurement_2010}. Another method to decouple spins from the electronic bath is to use a superconducting substrate. The energy gap around the Fermi level (E$_F$) can efficiently protect the spin excitations from decaying through generation of electron-hole pairs. For example, spin excitation lifetimes reaching $\approx$10~\si{\nano\second} have been observed in a Fe-based porphyrin over a lead substrate \cite{heinrich_protection_2013}. Unfortunately, many elemental
superconductors  strongly hybridize with molecular
species, quenching their spin state, or simply are difficult to integrate into large scale devices.  

Here, we took advantage of the proximity effect to induce a superconducting gap on a gold thin film \cite{Truscott1999,Gupta2004,Island2017,Wei2019} that extended the excitation lifetime of a tetraphenylporphyrin iron(III) chloride (FeTPP-Cl) for more than  80~\si{\nano\second}.   
High quality gold films were epitaxially grown on top of the oxygen reconstructed 1x5-V(100) surface. We used a scanning tunneling microscope (STM) at 1.2~\si{\kelvin} to perform Inelastic Electron Tunneling Spectroscopy (IETS) on FeTPP-Cl molecules adsorbed on the Au/V(100) surface and revealed the existence of such very long spin excitation lifetimes. Moreover, we found that increasing the thickness of the gold layer resulted in a reduction of the  excited state lifetime, which coincided with the gradual shift down of the proximity-induced in-gap resonances. By approaching the STM tip to the molecules, we modified the magnetic coupling with the substrate  to a stronger regime, where Yu-Shiba-Rusinov (YSR)\cite{yu_-_1965,shiba_classical_1968,rusinov} states appear inside the gap\cite{heinrich_single_2018}.


\begin{figure}[ht!]
	\includegraphics[width=0.9\columnwidth]{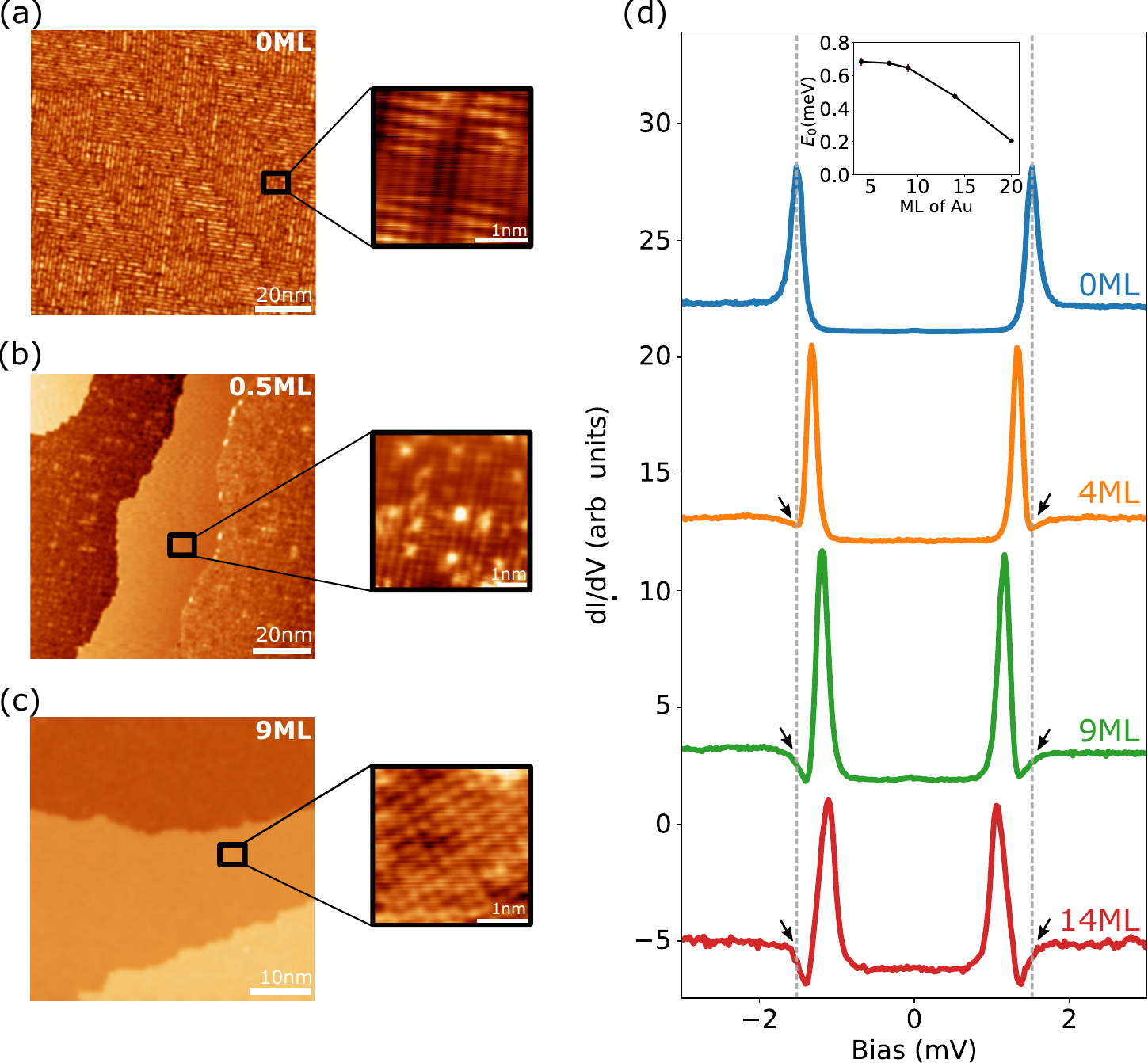}
	\caption{\label{fig:Fig1} \textbf{(a)} STM image of the oxygen reconstructed 5$\times$1-V(100) surface, where oxygen atoms occupy bridge positions along every fifth V-atom row (as shown in inset); I= 1~\si{\nano\ampere}, V=700~\si{\milli\volt}; inset: I= 3~\si{\nano\ampere}, V= 3~\si{\milli\volt}. 
	\textbf{(b)} STM image after depositing 0.5ML (I= 100~\si{\pico\ampere}, V= -300~\si{\milli\volt}; inset: I= 100~\si{\pico\ampere}, V= 5~\si{\milli\volt}) and \textbf{(c)} 9ML of Au on V(100)-O(5$\times$1) (I= 100~\si{\pico\ampere}, V= 1~\si{\volt}; inset: I= 400~\si{\pico\ampere}, V= 50~\si{\milli\volt}). \textbf{(d)}  STS spectra (using a superconducting V-covered tip) on different number of Au layers on V(100)-O(5$\times$1). Inset: evolution of the dGSJ intra-gap states compared to the amount of grown Au layers.}
\end{figure}

Vanadium is a type-II superconductor with a critical temperature of $T_{c}=5.4\si{\kelvin}$. Vanadium single crystals contain impurities embedded in the bulk, namely oxygen, carbon and nitrogen \cite{SuperconductivityofVanadium,VALLA1994843}. In particular, the presence of oxygen creates a reconstructed V(100)-O(5×1) surface \cite{vanadium-latticepar,PrepVanadium, vanadium-reconstruction, Bischoff2002ScanningTS} persisting even after very high temperature annealing (Figure~\ref{fig:Fig1}a). Scanning Tunneling Spectroscopy (STS) measurements were performed to study the magnitude of the superconducting gap. To maximize the energy resolution, the tip was indented in the V sample to become superconducting \cite{pan_vacuum_1998}. Differential conductance (dI/dV) spectra on the V(100)-O(5×1) surface revealed two clear dI/dV peaks at V=$\pm$1.5 \si{\milli\volt} (blue plot in Figure~\ref{fig:Fig1}d), corresponding to tunneling between coherence peaks of tip and sample at V = $\pm( \Delta_t+\Delta_s$), and in agreement with a  gap value for bulk  vanadium of $2\Delta_s\approx1.5~\si{\milli\electronvolt}$ \cite{huang_quantum_2020}. 

\begin{figure}[!th]
	\includegraphics[width=0.8\columnwidth]{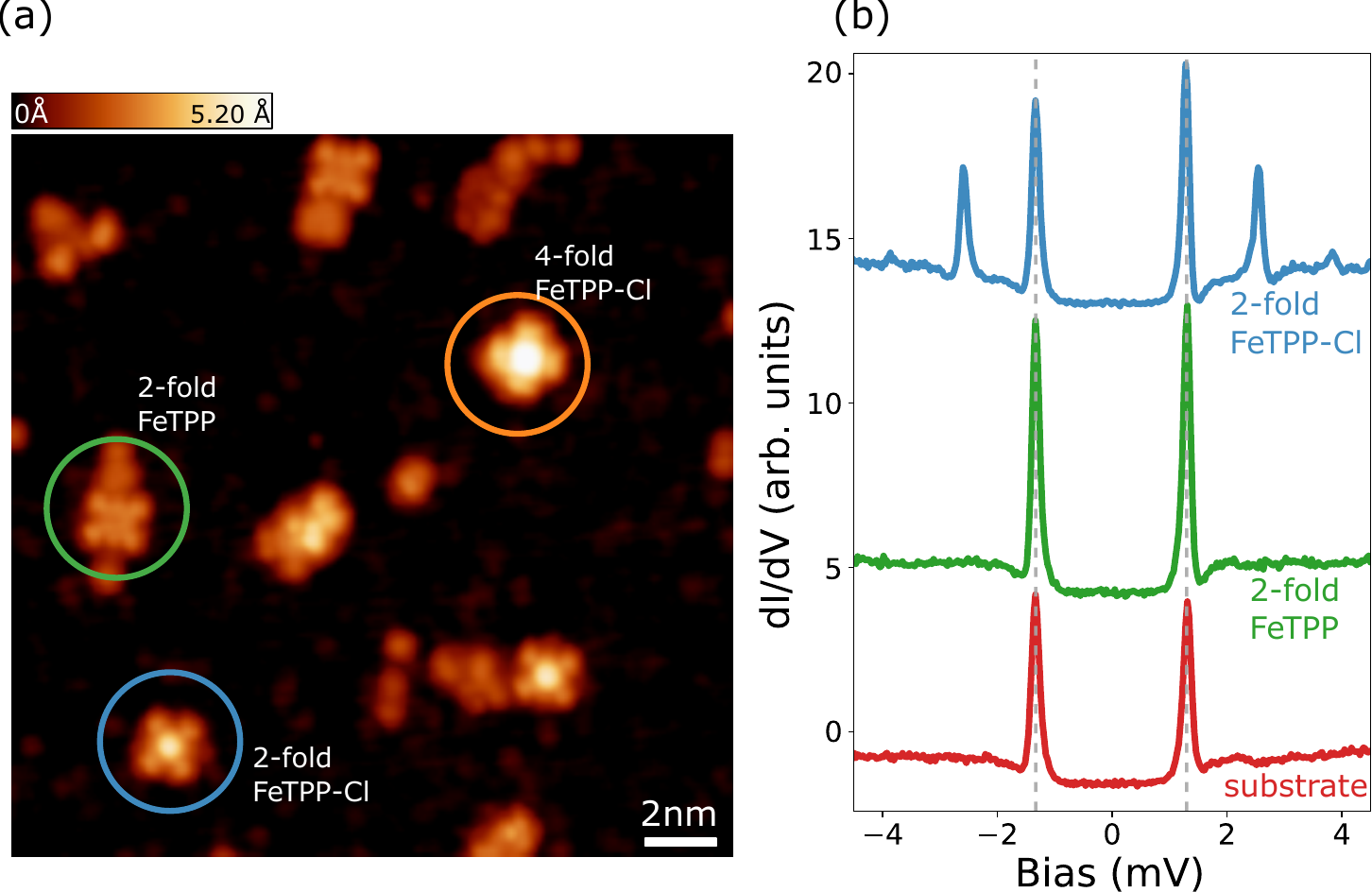}
	\caption{\label{fig:Fig2} 
	\textbf{(a)}: STM image of the 4-fold FeTPP-Cl (orange circle), 2-fold FeTPP-Cl (blue circle) and  2-fold FeTPP (green circle) on Au/V(100) (16x16 ~\si{\nano\meter}, I=20~\si{\pico\ampere}, V=200~\si{\milli\volt})  , \textbf{(b)}: STS spectra of the different molecular species on Au/V(100) (I=200~\si{\pico\ampere}, V=5~\si{\milli\volt})
	}
\end{figure}

Molecular adsorption on bare V(100) \cite{etzkorn2018mapping} is challenging due to the high reactivity of the surface, which can distort molecular structure and its functionality. Therefore, we covered the vanadium surface with gold films, a metal typically employed for hosting molecular assemblies. We grew multiple gold layers on the V(100)-O(5$\times$1) surface and studied the formation of superconducting pairing correlations through the proximity effect \cite{Island2017,Beck2022}. In Figure~\ref{fig:Fig1}b,c, STM images reveal the epitaxial growth of gold on the V(100)-O(5$\times$1) (herein referred to as V(100)) surface after annealing the substrate to 550$^{\circ}$C (see methods). Atomic resolution images of the gold lattice reveal the high quality of the films, with (100) surface orientation and lattice constant $x=y\approx3\si{\AA}$.

STS measurements on the gold films show that they maintain the superconducting character of the underlying V(100) crystal with a proximity-induced gap of width  $2\Delta_s\approx1.5~\si{\milli\electronvolt}$. For all film thicknesses explored in Figure~\ref{fig:Fig1}d, spectra show characteristic shoulders (black arrows in Figure~\ref{fig:Fig1}d) close to $\pm( \Delta_s+\Delta_t)$, the convoluted gap of the vanadium bulk and tip. This is expected for proximitized films thinner than the superconducting coherence length of vanadium \cite{arnold-theory,Truscott1999}. In addition, a pair of sharp spectral resonances appear inside the proximity gap. These resonances, first described by de~Gennes-Saint James \cite{deGennes1963}, are intra-gap  quasiparticle excitation peaks resulting from interfering pathways between the metal-vacuum surface and Andreev Reflections at the metal-superconductor interface \cite{arnold-theory,Wolf1982ProximityET}. The de Gennes-Saint James resonances (dGSJ) shift down in energy with increasing film thickness, from 0.69~\si{\milli\electronvolt} (4 ML) to 0.2~\si{\milli\electronvolt} (20 ML)  (see Supporting Information\cite{SI}).

Next, we deposited FeTPP-Cl molecules on the Au/V(100) system. As previously reported  \cite{gopakumar_transfer_2012,Heinrich2013}, deposition of FeTPP-Cl molecules on metallic substrates at room temperature usually results in coexistence of chlorinated and dechlorinated FeTPP species (Figure~\ref{fig:Fig2}a). The presence of the chlorine atom in STM topographic images appears as a protrusion in the center of the molecule, whereas the dechlorinated FeTPP molecules show a depression.
Furthermore, as shown in Figure~\ref{fig:Fig2}a, FeTPP-Cl appears with two distinctive adsorption geometries, referred to as 4-fold and 2-fold structures, due to the different apparent positions of the phenyl rings. The 4-fold species lie in a stronger interaction regime with the surface that will be described elsewhere. In contrast with previous results on Pb(111) and Au(111) \cite{Farinacci2018,Farinacci2020b,carmen,rubio-verdu_orbital-selective_2018,Carmen-thesis}, dechlorinated molecules on this substrate do not exhibit neither YSR in-gap states nor inelastic spin excitations outside the gap (Figure~\ref{fig:Fig2}b), indicating that they may lie in a different coupling regime.

\begin{figure*}[ht!]
	\includegraphics[width=1\textwidth]{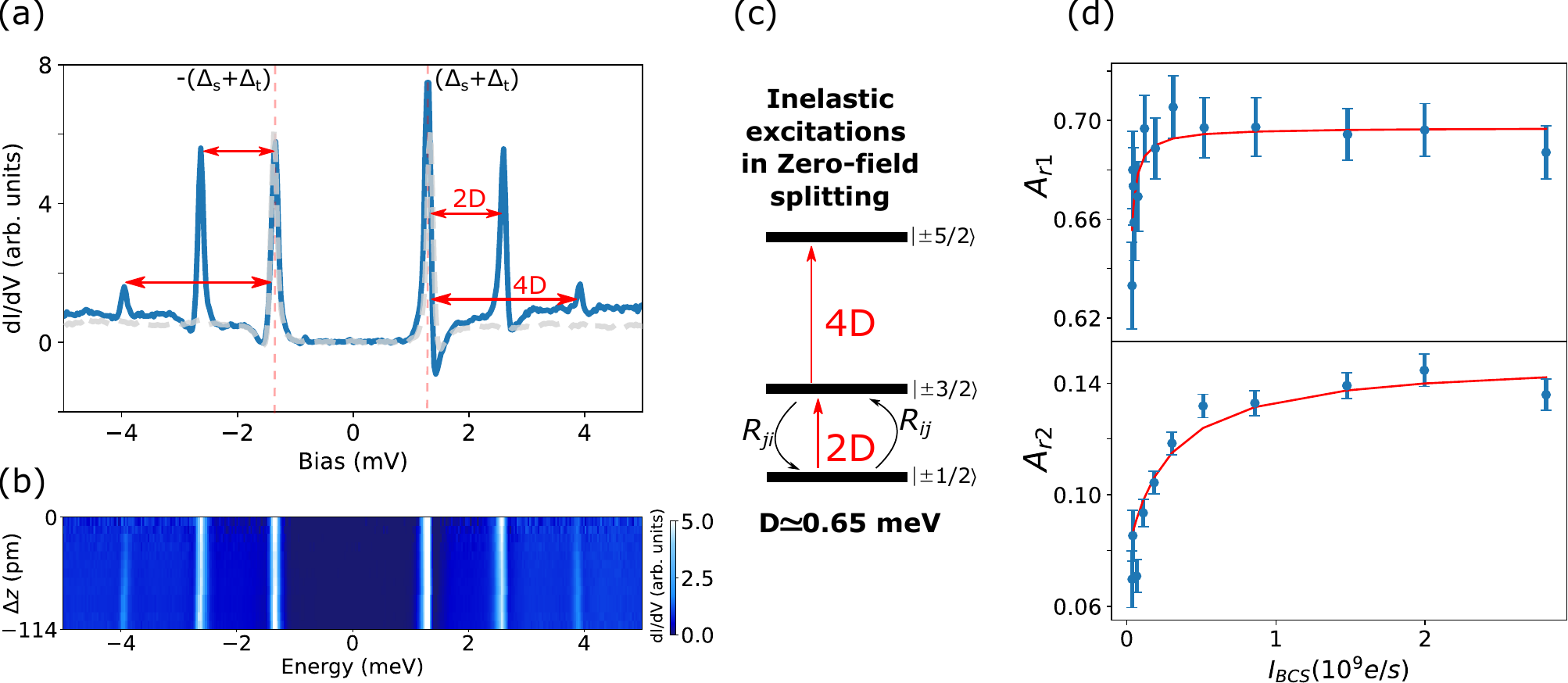}
	\caption{\label{fig:Fig3}  \textbf{(a)} Single STS spectra taken on a 2-fold FeTPP-Cl molecule (blue) and on the substrate (grey) (I=150~\si{\pico\ampere}, V=6~\si{\milli\volt}). \textbf{(b)} 2D intensity plot of a series of distance-dependent dI/dV spectra for 4~ML Au thickness. \textbf{(c)} Zero field splitting excitations for a total spin of $S_z=5/2$ ,
	 \textbf{(d)}{ Top (bottom) panel: Relative amplitude $A_{r1}$ ($A_{r2}$) of the first (second) excitation resonance as a function of $I_{BCS}$, the current of the BCS peaks. Error bars indicate the error propagation of the relative amplitude averaged over positive and negative side of the spectra. \textbf{Top}: Fit (red line) for determining the lifetime of the first excited state $\pm\ket{3/2}$ with thickness of 4ML ($P=0.423 \pm 0.001, \tau_{1}=80~\si{\nano\second} \pm 20~\si{\nano\second} $ ), \textbf{Bottom}: Fit (red line) for determining the lifetime of the second excited state $\pm\ket{5/2}$ with thickness of 4ML, with fixed $\tau_{1}$ ($ P = 0.304 \pm 0.014; \tau_{2\rightarrow1} = 1.8~\si{\nano\second} \pm 0.8~\si{\nano\second}$)}
	}
\end{figure*}

In this letter, we focus on the 2-fold FeTPP-Cl molecules. Characteristic dI/dV curves recorded on the molecular center show excitation peaks outside the superconducting gap, at $\approx \pm 2.6 ~\si{\milli\volt}$ and $\approx \pm 3.8 ~\si{\milli\volt}$ (Figure~\ref{fig:Fig3}a). These peaks have been previously attributed to inelastic excitations of the $S=5/2$ molecular spin multiplet\cite{heinrich_protection_2013,theoryIETS}. 
Using the phenomenological hamiltonian $H=DS^2_{z}$ to describe the effect of a ligand field, a positive magnetic uniaxial anisotropy $D$ found for these molecules splits the degenerate Fe multiplet into three spin states, $S_z=\pm\ket{1/2},\pm\ket{3/2},\pm\ket{5/2}$  \cite{gatteschi_molecular_2006}. 
The inelastic signal appears as replica of the dGSJ peaks in the spectra and the energy separation between them (red arrows in Figure~\ref{fig:Fig3}a) corresponds to the excitation energies. That is, 2$D$ from the ground state $S_z=\pm\ket{1/2}$ to the first excited state $S_z=\pm\ket{3/2}$, and 4$D$ subsequently to the second excited state, $S_z=\pm\ket{5/2}$ (Figure~\ref{fig:Fig3}c). Therefore, from these measurements we obtain that the axial magnetic anisotropy for FeTPP-Cl amounts to $D=0.65~\si{\milli\electronvolt}$. 
Due to the spin selection rule $\Delta S_z=\pm1$, the occupation of the second excited state can only be achieved from the first excited state \cite{loth_controlling_2010,heinrich_protection_2013}. The observation of a peak at 4$D$ thus indicates that the first excited state reaches a finite population in a  stationary state with the tunneling current \cite{theoryIETS,heinrich_protection_2013}. Such step-wise excitation of higher lying spin states is referred to as spin-pumping \cite{loth_controlling_2010}.

We study the spin excitation dynamics of 2-fold molecules first on a 4 monolayer (ML) thick Au film on V(100), with characteristic spectra like the one shown in Figure~\ref{fig:Fig3}a. Interestingly, the excitation into the 
$S_z=\pm\ket{5/2}$ state appears for tunneling currents as small as 20 \si{\pico\ampere}, suggesting a particularly long lifetime for  the first excited state ($S_z=\pm\ket{3/2}$). The lifetime of the excited states, $\tau_1$, can be extracted from the behaviour of the  IETS peaks with current, using a set of rate equations $\frac{dN_i}{dt}=\sum\limits_{j}(R_{ji}N_j-R_{ij}N_i)$ connecting states $N_i$ and $N_j$ with excitation/de-excitation rates $R_{ij}$ and $R_{ji}$ \cite{heinrich_protection_2013}. 

For the first excited state, the de-excitation rate $R_{ji}$ includes a spontaneous decay constant, $\lambda_{1}$, and a current-induced de-excitation factor $f$ (see Supporting Information \cite{SI}). 
The former gives the lifetime of the first excited state ($\lambda_{1} = 1 / \tau_{1}$) and defines the threshold to a stationary non-equilibrium state. In this regime, only the inelastic current drives transitions between the ground state and the first excited state, which dynamically acquires a finite population. This opens a new conduction channel for the second excited state.

To determine $\tau_{1}$, we acquired multiple spectra similar to the one shown in Figure~\ref{fig:Fig3}a with increasing set-point current. A stacked plot  of distance-dependent dI/dV spectra is shown in Figure~\ref{fig:Fig3}b, where we observe an increase in the intensity of the excitation peaks as the tip approaches the molecule. The ratio $A_{r1}$ between the amplitude of the first excitation peak, $A_{1}$, and that of the coherence peaks, $A_{BCS}$, is plotted in Figure~\ref{fig:Fig3}d (top panel) as a function of $I_{BCS}$, the current at the coherence peaks. The inelastic amplitude first increases with the tunneling rate and then stabilizes in a plateau characteristic of a stationary state. 
Fitting these data with rate equations from \citet{heinrich_protection_2013} yields a value of $\tau_{1}\approx 80~\si{\nano\second}$ for FeTPP-Cl on 4ML of Au on V(100) (see fitting procedures in the Supporting Information \cite{SI}). Such large lifetime is attributed to the protection of the excited state by the superconducting gap, which forbids spin relaxation via creation of electron-hole excitations because $2D<2\Delta_s$. 

Interestingly, the value of $\tau_{1}$ obtained for FeTPP-Cl on Au/V(100) is 8 times larger than the related molecule Fe-OEP-Cl on Pb(111) \cite{heinrich_protection_2013}, where the de-excitation is protected by an even larger superconducting gap. As previously shown  \cite{electron_spin_lifetime_limited_by_phononic_vacuum_modes,relax-mechanisms}, another decaying mechanism could take place through the direct emission of phonons, with energy $\omega\hbar \approx 2D$, to the phononic bath of the substrate. Thus, a possible origin for the larger lifetime on this substrate could be the lower phonon density of
states at the Fermi level of vanadium and gold compared to lead \cite{Electron-phonon-DOS-vanadium, Au-phon-DOS,Phonon_Spectra_in_Pb}, as well as the lower electron-phonon coupling present in Au \cite{McMillan_e_ph_constant,Au-phonon}. From the stacked spectra in Figure~\ref{fig:Fig3}b, we can also obtain an estimation of the lifetime $\tau_{2\rightarrow1}$ of the second excited state $S_z=\pm\ket{5/2}$.  
In the bottom panel of Figure~\ref{fig:Fig3}d, we plot the relative amplitude of the second excitation peak, $A_{r2}$, with respect to $I_{BCS}$. Fitting the data with a set of rate equations (red line) yields a value of $\tau_{2\rightarrow1}\lesssim 2~\si{\nano\second}$ (see Supporting Information~\cite{SI}). Electrons populating the second excited state have now sufficient energy ($4D>2\Delta_s$) to directly decay into the substrate by electron-hole pair excitations, lowering the lifetime of the highest excited state. Nevertheless, this value is still relative large, and comparable to magnetic atoms on insulating films \cite{loth_controlling_2010}.

\begin{figure*}[!t]
	\includegraphics[width=1\textwidth]{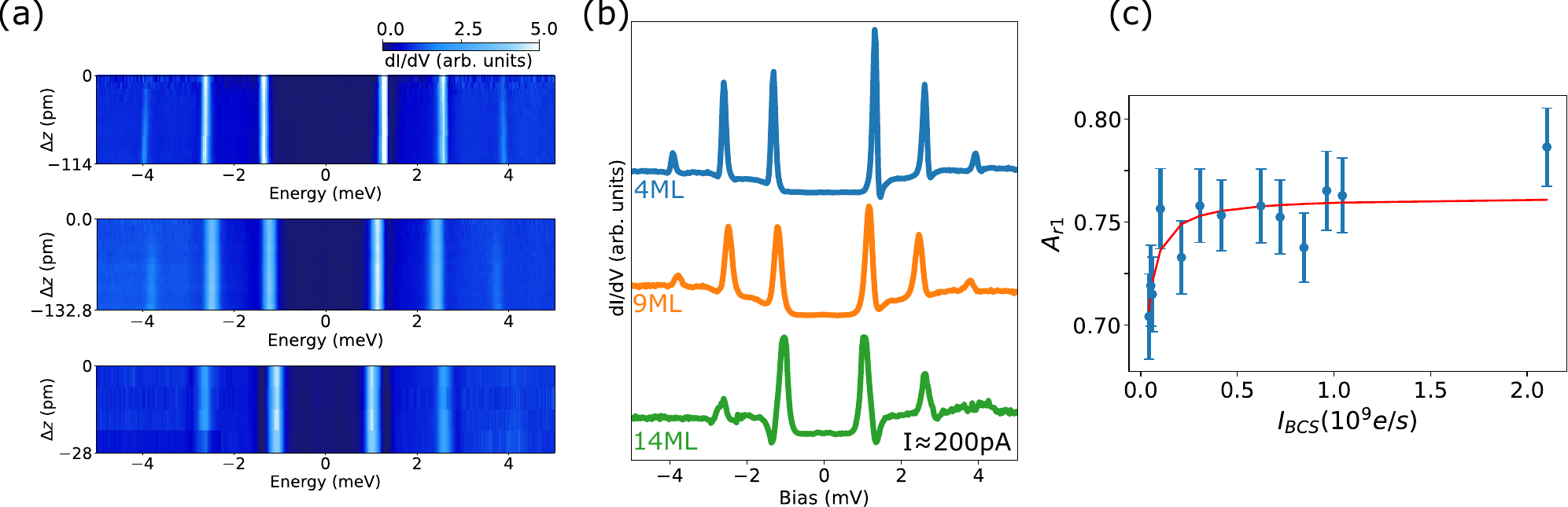}
	\caption{\label{fig:Fig4} \textbf{(a)} 2D intensity plots of a series of distance-dependent dI/dV spectra for 4~ML, 9ML and 14ML Au thickness on V(100). \textbf{(b)} Single STS spectra taken on a FeTPP-Cl molecule on 4ML (blue), 9ML (orange) and 14ML (green) Au thicknesses on V(100). (I=200~\si{\pico\ampere}, V=6~\si{\milli\volt}). \textbf{(c)}: Fit for determining the lifetime of the first excited state $\pm\ket{3/2}$ with thickness of 9ML ($P = 0.460 \pm 0.003, \tau_{1} = 39~\si{\nano\second} \pm 10~\si{\nano\second} $)}
\end{figure*}

To test the influence of the intra-gap bound states on the lifetimes of the excited states, we compared FeTPP-Cl spin excitation dynamics in samples with different Au thicknesses. In Figure~\ref{fig:Fig4}a, two dimensional intensity plots of a series of distance-dependent dI/dV spectra for 4, 9 and 14~ML Au thickness are shown. The plots show the shift of the dGSJ resonances to lower energies as a function of the Au thickness, similar to Figure~\ref{fig:Fig1}b. 

We follow the same fitting procedure for determining the excited state lifetime of FeTPP-Cl on every film thickness. For 9~ML, the fit yields a value of $\tau_{1}=39~\si{\nano\second} \pm 10~\si{\nano\second}$ and $\tau_{2\rightarrow1}=1.1~\si{\nano\second}\pm2.2~\si{\nano\second}$  for the lifetime of the first and second excited state, respectively (Figure~\ref{fig:Fig4}c and Figure~S2).
On the 14~ML film, we could hardly observe faint second excitation peaks (Figure~\ref{fig:Fig4}b) by approaching the tip to the closest possible position ($I=200~\si{\pico\ampere}$) before picking up the molecule. This determined an upper limit for the spin excitation of $\tau_{1}<1~\si{\nano\second}$. 

The reduction in excitation lifetime is in agreement with the gradual shift  of the dGSJ resonances in the proximitized Au film. From Figure~\ref{fig:Fig1}d, we obtain that the bound state values decrease from $\sim0.69 \si{\milli\electronvolt}$ (4~ML) to $\sim0.65 \si{\milli\electronvolt}$ (9~ML), and finally to $\sim0.45 \si{\milli\electronvolt}$ (14~ML). In contrast, the magnetic anisotropy, as obtained from the position of the spectral peaks, remains $D\approx0.65~\si{\milli\electronvolt}$ for the 4 and 9~ML of Au on V(100), while increases to $D\approx0.8~\si{\milli\electronvolt}$ for the 14ML film. Variations in $D$ have been previously observed in different magnetic systems on metallic substrates and attributed to either structural changes \cite{heinrich_protection_2013,heinrich_tuning_2015,kezilebieke_observation_2019} or variations in the Kondo coupling with the substrate \cite{oberg_control_2014}. Comparing the spectra of FeTPP-Cl on the different gold thicknesses in Figure~\ref{fig:Fig4}b, we note that the dGSJ  peaks show a clear asymmetry for the 4~ML sample, that decreases for 9~ML, and vanishes for the 14~ML case.
Such spectral imbalance reflects a particle-hole asymmetry in the quasi-particle excitations of the superconductor probably due to the presence of a finite coupling with the molecule. The absence of peak asymmetry for the 14~ML reveals a significant smaller Kondo coupling with the substrate, which can explain a renormalized magnetic anisotropy to a larger value \cite{delgado_consequences_2014, jacob_renormalization_2018}.

 \begin{figure}[th!]
	\includegraphics[width=0.8\columnwidth]{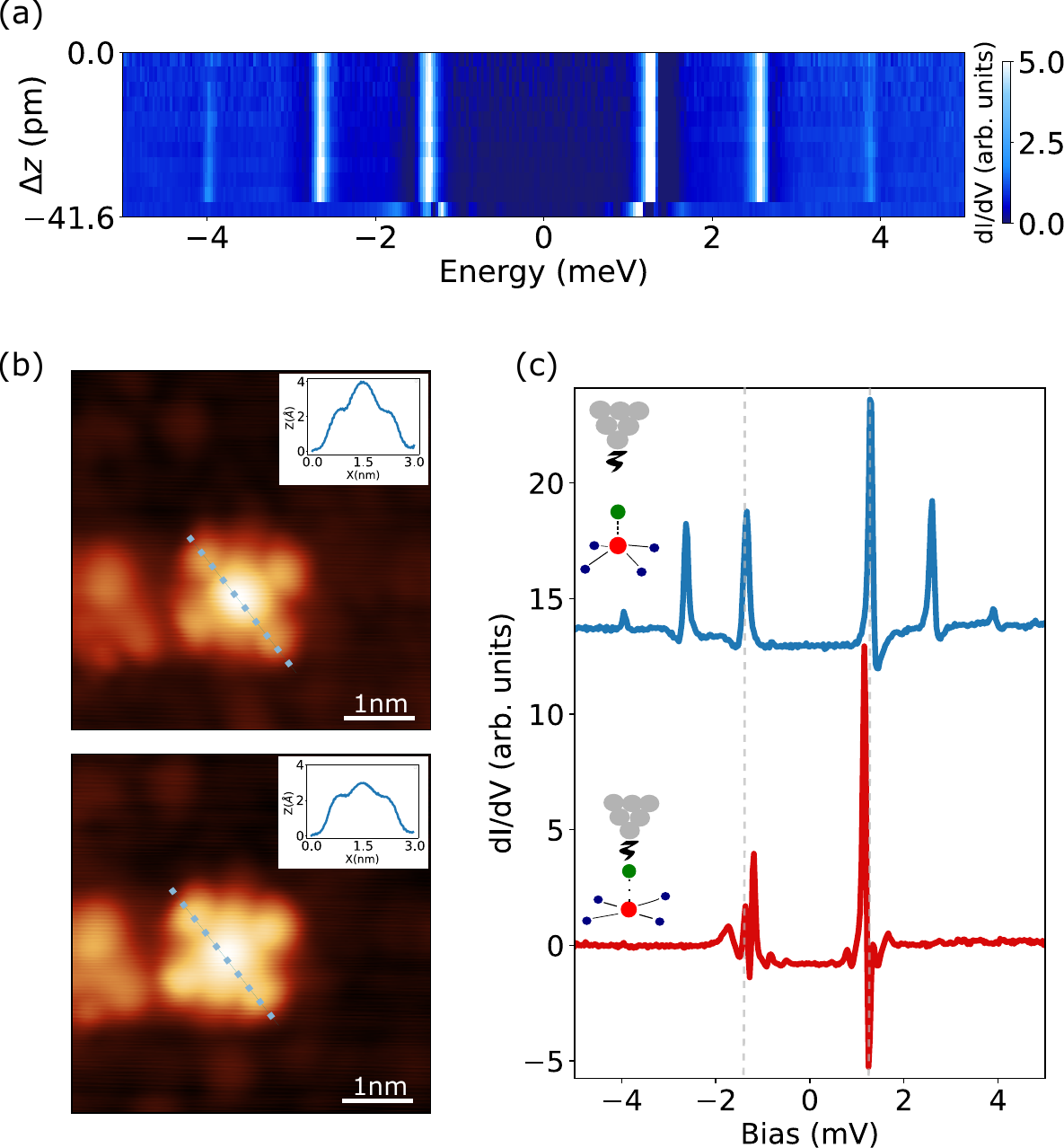}
	\caption{\label{fig:Fig5} \textbf{(a)} 2D intensity plot of a series of distance-dependent dI/dV spectra for a switching FeTPP-Cl molecule. \textbf{(b)} STM images taken before (top) and after (bottom) the change of the interaction regime. Insets: line profiles taken across the molecule as indicated by the blue dotted line. \textbf{(c)} Single dI/dV spectra before (top) and after (bottom) the change of the interaction regime. Insets: Cartoon model representing the tip-induced effect far from (top) and close to (bottom) the molecule.}
\end{figure}

The weak adsorption regime of this molecular species could be irreversibly modified by approaching the tip beyond a specific distance. In this case, we observed an abrupt change in the dI/dV signal (Figure~\ref{fig:Fig5}a). The two characteristic spin excitation peaks disappear, while two pairs of YSR states arise within the superconducting gap (Figure~\ref{fig:Fig5}c, bottom spectra). The emergence of YSR states indicates the increase of the exchange coupling strength, $J$, of the molecular spin with the Cooper pairs of the superconducting condensate  \cite{heinrich_single_2018}. 
Such a change in the interaction regime,  previously observed by~\citet{heinrich_control_2018},  is understood as a slight variation of the molecular geometry induced by the tip in close proximity to the molecule. Attractive forces exerted to the Cl atom, weaken the Cl-Fe bond and Fe is allowed to relax in the molecular plane as shown in the cartoon of Figure~\ref{fig:Fig5}c.  
In order to verify that this change of coupling regime is not due to the detachment of the Cl ligand, or any other irreversible molecular distortion, we checked the STM images before and after the tip approach (Figure~\ref{fig:Fig5}b). Additionally, a broad peak emerges in the new regime outside the superconducting gap at $eV\approx1.7 ~\si{\milli\volt}$. 
This new molecular configuration has been associated with a different oxidation state,  with $S=2$, stabilized by the hybridization of the Fe ion with the surface \cite{uzma_understanding_2021}.

In summary, in this work we have shown that proximity-induced superconducting gold thin films can enhance the spin excitation lifetime of magnetic molecules. We explored the spin excitation lifetimes of an S=5/2 iron(III) porphyrin-chloride molecule on Au films grown on V(100) by implementing IETS with an STM. Our results found spin excitation lifetimes of $\approx80 \si{\nano\second}$ for the excitation of the  S$_z=\pm$3/2 multiplet, which is protected from decaying to an electron-hole pair excitation by the absolute proximity-induced gap of the substrate. 
The excitation lifetime in this surface platform is even larger than the one found for related species on Pb(111) \cite{heinrich_protection_2013}. This is attributed to the lower density of phononic states in the substrate, suggesting that direct phonon emission is a lifetime limiting process. However, we also determined that the lifetime is affected by a pair of in-gap de Gennes-Saint James resonances in the film, which separate from the gap edge and shift inwards towards zero energy, thus opening alternative (coherent) channels for decaying. Overall, the Au/V(100) platform offers the advantage of employing a noble metal, mild in molecular adsorption and reactions,  for studies of long spin dynamics. Additionally,  the presence of coherent sub-gap states in the proximitized metal film suggests new potential methods for addressing spin states and excitations of magnetic molecules.

\section{Methods}
All experiments were carried out on a commercial ultra-high vacuum scanning tunneling microscope of SPECS with a Joule-Thomson cooling stage reaching a base temperature of $1.2~\si{\kelvin}$. For the acquired STS spectra, an external lock-in was used with a modulation frequency of $938~\si{\hertz}$ and a amplitude of $V_{rms}=20-50~\si{\micro\volt}$. 

The vanadium (100) crystal used in this project was repeatedly sputtered with Ar$^{+}$ at 1.5\si{\kilo\volt} in UHV conditions ($P=5\times10^{-9}\si{\milli\bar}$) and annealed at high temperatures ($\sim 1000^{\circ}$C) in an attempt to remove the residual impurities as suggested by Kralj et al. \cite{PrepVanadium}. Gold thin films were deposited on V(100) with an e-beam evaporator, while maintaining the crystal at 550$^{\circ}$C. Afterwards, the samples were post-annealed for 10 minutes at 600 $^{\circ}$C. The calibration of the deposition rate was made by evaporating 0.5 monolayers and subsequently checking the Au coverage by STM.

The tip was prepared by indenting up to 20$\si{\nano\meter}$ into the substrate. A superconducting tip results in a substantial increase of the energy resolution, since tunneling of electrons takes place between the two sharp DOS of tip and sample \cite{pan_vacuum_1998}. Sublimation of FeTPP-Cl molecules from a Knudsen cell at 297$^{\circ}$C resulted in a mixture of FeTPP-Cl and FeTPP species on the surface, shown in Figure~\ref{fig:Fig2}.  

\begin{suppinfo}
Deconvolution of acquired dI/dV spectra for different Au thicknesses on V(100), as well as fit for the energies of the intra-gap states based on \citet{deGennes1963} model are provided. A brief description of the rate equations leading to the fitting function for $A_{r1}$ and $A_{r2}$ and the fit for extracting the lifetime of the second excited state of the sample with 9ML of Au are presented. Finally, a magnetic field dependent measurement is shown.  
\end{suppinfo}

\begin{acknowledgement}
The authors thank Sebastian Bergeret for stimulating and helpful discussions, and Christian Ast for information regarding the preparation of the V(100) surface.  We also gratefully acknowledge financial support from the \textit{Red guipuzcoana de Ciencia, Tecnología e Innovación} through program NEXT01, the Agencia Estatal de Investigaci\'{o}n (AEI) (project No. FIS2017-83780-C1 and the Maria de Maeztu Units of Excellence Programme MDM-2016-0618), the European Regional Development Fund, and the European Union (EU) H2020 program through the FET Open project SPRING (grant agreement No.~863098). J.O. thanks the Basque Departamento de Educaci\'on through the PhD scholarship no.~PRE\_2019\_2\_0218.

 \end{acknowledgement}

\bibliographystyle{apsrev4-1}


\providecommand{\latin}[1]{#1}
\makeatletter
\providecommand{\doi}
  {\begingroup\let\do\@makeother\dospecials
  \catcode`\{=1 \catcode`\}=2 \doi@aux}
\providecommand{\doi@aux}[1]{\endgroup\texttt{#1}}
\makeatother
\providecommand*\mcitethebibliography{\thebibliography}
\csname @ifundefined\endcsname{endmcitethebibliography}
  {\let\endmcitethebibliography\endthebibliography}{}

\end{document}


\title{Extending the spin excitation lifetime of a magnetic molecule on a proximitized superconductor}

\author{Katerina Vaxevani}
  \affiliation{CIC nanoGUNE-BRTA, 20018 Donostia-San Sebasti\'an, Spain}
  
\author{Jingcheng Li}
  \affiliation{CIC nanoGUNE-BRTA, 20018 Donostia-San Sebasti\'an, Spain}
\affiliation{School of Physics, Sun Yat-sen University, Guangzhou 510275, China}
\author{Stefano Trivini}
  \affiliation{CIC nanoGUNE-BRTA, 20018 Donostia-San Sebasti\'an, Spain}
  
  \author{Jon Ortuzar}  
  \affiliation{CIC nanoGUNE-BRTA, 20018 Donostia-San Sebasti\'an, Spain}
  
\author{Danilo Longo}  
  \affiliation{CIC nanoGUNE-BRTA, 20018 Donostia-San Sebasti\'an, Spain}
  
\author{Dongfei Wang}  
  \affiliation{CIC nanoGUNE-BRTA, 20018 Donostia-San Sebasti\'an, Spain}
  
\author{Jose Ignacio Pascual}
  \affiliation{CIC nanoGUNE-BRTA, 20018 Donostia-San Sebasti\'an, Spain}
  
\affiliation{Ikerbasque, Basque Foundation for Science, 48013 Bilbao, Spain}\email{ji.pascual@nanogune.eu}

\date{\today}
\renewcommand{\abstractname}{\vspace{4cm} }	
\begin{abstract}
\baselineskip12pt
	\tableofcontents		
\end{abstract}	
\baselineskip14pt
\maketitle

\setstretch{1.0}

\setstretch{1.1}

\section{Shift of \textit{de Gennes-Saint James} resonances with gold film thickness}

As shown in Figure~1d of the main manuscript, the pair of the intra-gap dGSJ resonances split and shift from the superconducting gap edge $\Delta_s$ with film thickness $d$. The evolution of dGSJ resonances' energy $E_{0}$ with $d$ in tunneling experiments has been described by Arnold and Wolf \cite{arnold-theory,Wolf1982ProximityET} through the expression: 
\begin{equation}
    E_{0} \simeq \Delta_s[1-\frac{1}{2}(R_{eff}d\Delta_s)^2]
\label{eqn:E0}
\end{equation}
where  $R_{eff}$ is a phenomenological parameter dependent on the Fermi velocity of the normal metal, as well as the electron reflection at the superconductor/normal metal (SN) interface. For clean and transparent interfaces, $R_{eff}$ is generally small and the dGSJ resonances appear with maximum energy close to $\Delta_s$ for large film thicknesses. 
For increasing reflection coefficient at the SN interface, $r\rightarrow1$, the effective thickness of the normal metal is increased by a factor $\frac{1+r}{1-r}$ and the resonances shift faster with $d$ \cite{deGennes1963,Wolf1982ProximityET,BARSAGI197729}. As shown in Supplementary Figure~\ref{SF2}, the dGSJ resonances in our experiment shift substantially for only 4\si{\nano\meter} thick films. Fitting eq.~\ref{eqn:E0} we obtain a value of $R_{eff}$ one order of magnitude larger than expected for the ideal case of a perfectly transmitting interface (r=0). This can be attributed to both a mismatch in Fermi velocities of Au(100) and V(100) and to the additional scattering at the Au/V interface.  

\begin{figure}[bh]
	\includegraphics[width=0.5\columnwidth]{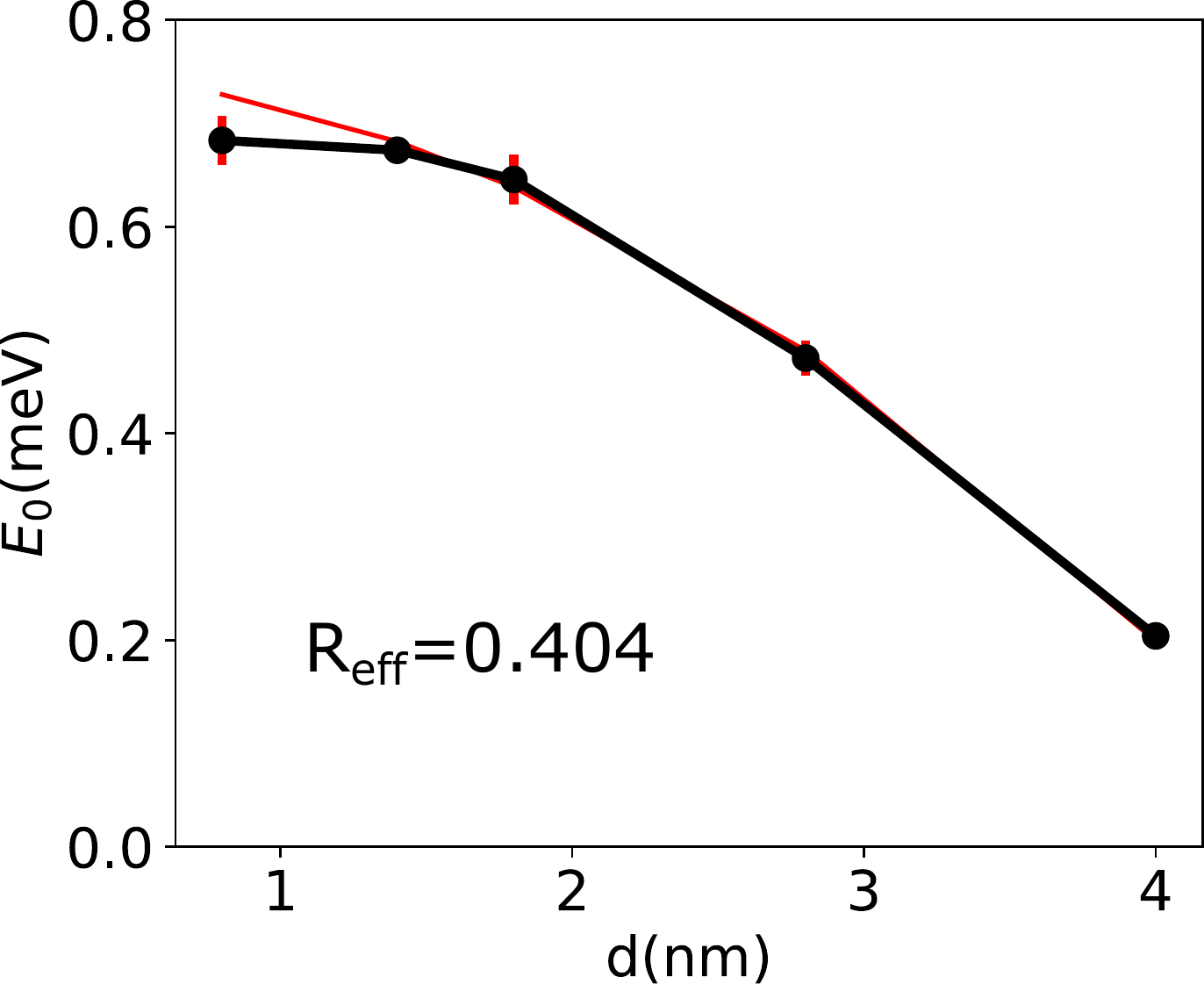}
	\caption{\label{fig:Fig3}Fitting of the evolution of the obtained values for the intra-gap dGSJ resonances with the sample Au thickness
	}\label{SF2}
\end{figure}

\newpage

\section{Rate equations for determining spin dynamics of excited states} 

Here we describe fitting functions obtained from rate equations in \cite{heinrich_protection_2013}. To extract the lifetime of the first excited state, $\tau_{1}$, we fit the inelastic amplitude plotted in the main manuscript, e.g. in Figure~3d (top),  with the expression:  

\begin{equation}
A_{r1}=P \frac{(I_{BCS}/e)P(2+2\epsilon_{1}/\Delta)+\lambda_{1}}{(I_{BCS}/e)P(2+\epsilon_{1}/\Delta)+\lambda_{1}}
\label{eqn:Ar1}
\end{equation}
where $P$ is the transition probability to the first excited state, $I_{BCS}$ is the elastic current at the onset of the dGSJ coherence peaks, $\lambda_{1}=1/\tau_{1}$ is the decay constant, $\epsilon_{1}$ is the energy of the excitation peak and $\Delta$ is here the energy of dGSJ states (we thus assume here for simplicity that the substrate density of states follows a BCS function with gap $\Delta$). The factor $f=(2+2\epsilon_{1}/\Delta)$ expresses the current-induced de-excitation processes by tunneling electrons.

The fitting function used to extract the lifetime of the second excited state, $\tau_{2}$, is
\begin{equation}
\resizebox{.9\hsize}{!}{$A_{r2}=P \frac{P^2 \frac{I_{BCS}}{e} (1 + \frac{\epsilon_{2} - \epsilon_{1}}{2\Delta}) \left[2P \frac{I_{BCS}}{e} (1 + \frac{\epsilon_{2}}{\Delta}) + \lambda_{2\rightarrow1} + \lambda_{2\rightarrow0}\right]}{\lambda_{1}(\lambda_{2\rightarrow1}+\lambda_{2\rightarrow0})+P \frac{I_{BCS}}{e}\left[\lambda_1(1+\frac{\epsilon_{2}}{\Delta})+\lambda_{2\rightarrow1}(2+\frac{\epsilon_{2}}{\Delta})+\lambda_{2\rightarrow0}(3+\frac{\epsilon_{2}}{\Delta})\right]+P^2 (\frac{I_{BCS}}{e})^2\left[3+\frac{7\epsilon_{2}}{2\Delta}-\frac{\epsilon_{1}}{2\Delta}+\frac{\epsilon^{2}_{2}}{\Delta^2}\right]}$}
\label{eqn:Ar2}
\end{equation}
where $P$ accounts here for both transitions to  excited states, $\lambda_{2\rightarrow1}=1/\tau_{2\rightarrow1}$ is the decay constant from the second to the first excited state and $\lambda_{2\rightarrow0}=1/\tau_{2\rightarrow0}$ to the ground state, and $\epsilon_{1}$ and $\epsilon_{2}$ are the energies of the first and second excitation peak. The lifetime $\tau_{2}$ of the second excited state can be calculated from
\begin{equation}
    \tau_{2} = \frac{1}{ \lambda_{2\rightarrow1} + \lambda_{2\rightarrow0}}.
\end{equation}
where the lifetime, $\tau_{2\rightarrow0}$, was fixed at 100\si{\nano\second} considering that this is a  phonon-mediated transition \cite{Leuenberger_1999} and that tunneling electrons do not contribute to it. 
The fittings from eq.\ref{eqn:Ar2} are presented in Figure~3d in the main text for the 4 monolayers and in Supplementary Figure S2 for 9 monolayers of the gold film.

\begin{figure}[bh]
	\includegraphics[width=0.4\columnwidth]{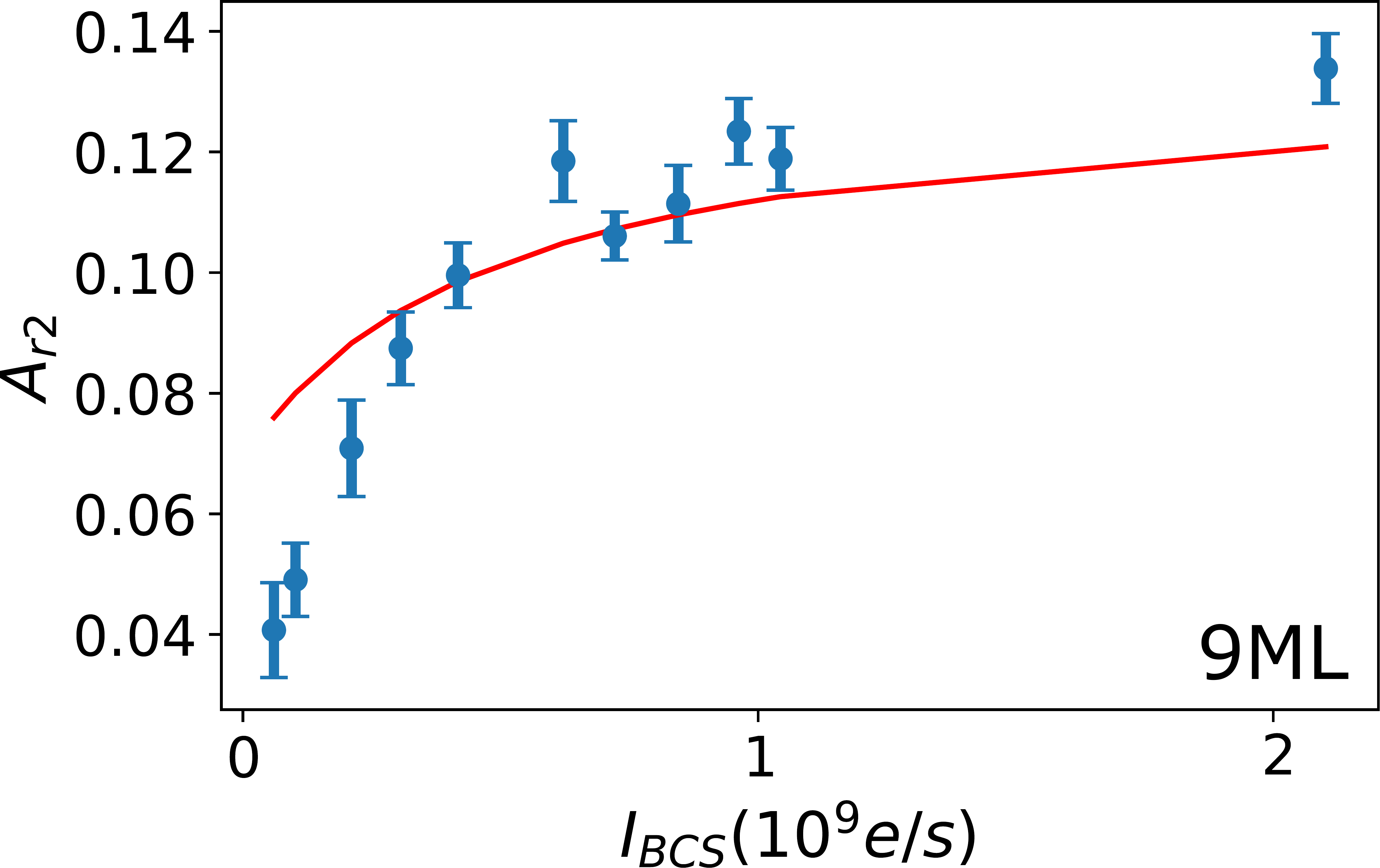}
	\caption{\label{fig:Fig2} Fit for determining the lifetime of the second excited state $\pm\ket{5/2}$ with Au thickness of 9ML with fixed $\tau_{1} = 39\si{\nano\second} \pm10\si{\nano\second}$, as obtained from fits to the first excitation peak. ($ P = 0.269 \pm 0.065 $ and $\tau_{2\rightarrow1} = 1.1\si{\nano\second} \pm 2.2\si{\nano\second}$).   
	}
\end{figure}

\clearpage

\section{Inelastic spectra in the quenched superconductor by a magnetic field}

In this section we present STS data taken on FeTPP-Cl with a magnetic field of 2.7\si{\tesla} applied perpendicular to the V(100) surface. The superconducting state of tip and sample are quenched and the inelastic spin excitations appear as steps in the dI/dV spectra. The normal state STS serves as a proof of the protection of the lifetime of the first excited state  by the superconducting gap, since no inelastic step for the second excitation can be seen. The plot was fitted using the code by M. Ternes\cite{ternes_spin_2015} to confirm that the dip shown here can be explained by the multiple transitions between the now Zeeman-split ground states ($S_z=+\ket{1/2}$ and $S_z=-\ket{1/2}$) as well as to the higher excited states ($S_z=+\ket{3/2}$ and $S_z=-\ket{3/2}$), with the same axial anisotropy $D=0.65$ meV. From the fit we find an effective temperature of 1.7\si{\kelvin}.   

\begin{figure}[th]
	\includegraphics[width=0.5\columnwidth]{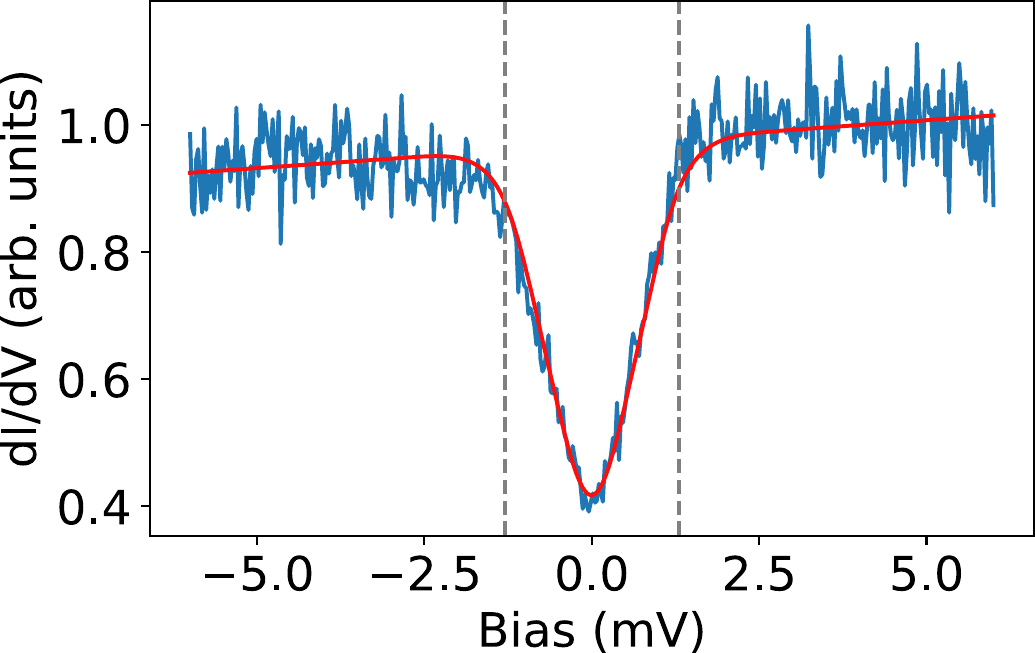}
	\caption{\label{fig:Fig4}STS on 2-fold FeTPP-Cl with applied magnetic field B=2.7\si{\tesla} (V=6\si{\milli\volt}, I=200\si{\pico\ampere}, $V_{rms}=0.5\si{\milli\volt}$) (blue line) and fitting (red line) with the 3rd order simulator from \citet{ternes_spin_2015}. 
	}
\end{figure}

\bibliographystyle{apsrev4-1} 

%
